\documentclass[preprint,12pt]{elsarticle}

\usepackage{amssymb}
\usepackage{amsmath}
\usepackage{amsfonts}

\journal{Nuclear Physics A}

\begin{document}

\begin{frontmatter}



\title{The lowest excited states of $^{14}$C and $^{14}$O nuclei within a
five-cluster model.}

\author[a1]{B. E. Grinyuk\corref{cor1}}
\ead{bgrinyuk@bitp.kiev.ua}\cortext[cor1]{}

\address[a1]{Bogolyubov Institute for Theoretical Physics,\\
 Kiev 03143, Ukraine}

\author[a1]{D. V. Piatnytskyi}
\ead{dvpyat@gmail.com}

\author[a1]{V. S. Vasilevsky}
\ead{vsvasilevsky@gmail.com}



\date{\today }
\begin{abstract}
We study the ground and the first excited $0^{+}$ states of two
mirror nuclei $^{14}$C and $^{14}$O within a five-cluster model
(three alpha-particles and two extra nucleons) with the use of high accuracy variational approach with Gaussian bases. The first excited $0^{+}$  state of these nuclei is shown to be connected with a change of the structure of the two-nucleon subsystem moving in
the field of the $^{12}$C cluster. The density distributions, electric form factors (both the elastic and transition ones), pair correlation functions, and the momentum distributions of the constituent particles are found. The most probable shapes of five-cluster systems are revealed.
\end{abstract}




\end{frontmatter}



\section{Introduction}

Cluster models are powerful tools to study a large verity of atomic nuclei and
nuclear reactions. They constantly demonstrate their ability to describe and
to interpret numerous phenomena in atomic nuclei observed both in scientific
laboratories and the Universe. They also allowed one to predict interesting
properties of nuclei or nuclear reactions. New developments of cluster models,
their effectiveness and new results obtained are thoroughly discussed in the
following reviews  \cite{2010RPPh...73c6301D, 2012PThPS.192....1H, 
 2015PrPNP..82...78F, 2016arXiv160302574B, 2018arXiv181208013B, 2018RvMP...90c5004F,
 2020FrPhy..1514401Z, 2020PTEP.2020lA101M}.

To provide more adequate description of nuclear structure and
nuclear processes, cluster models are subject for a permanent
improvement and advancing, which, for instance, include more and
more clusters and application to study nuclei with a large excess
of protons or neutrons. By including more clusters and
consequently more channels into consideration one faces with bulky
calculations, which are also complicated by the Pauli principle.
There were formulated some methods to reduce calculations
associated with the Pauli principle. One of the most famous
methods is the orthogonality condition method suggested by S.
Saito \cite{Saito69,kn:Saito77}.
This method significantly reduces computational efforts and
provides a reasonable description of nuclear structure and
reactions. There are also further simplifications when clusters are
considered to be structureless particles and cluster-cluster
interactions are modelled by some effective interactions. Such a
type of cluster models does not require tremendous calculations
and allows one to reveal some interesting phenomena which can be
confirmed in more sophisticated calculations.

We consider the structure of $^{14}$C and $^{14}$O within a
five-body model. These nuclei are presented as a
configuration $\alpha+\alpha +\alpha+N+N$. This five-cluster
configuration allows one, in principle, to analyze the structure of $^{14}$C and
$^{14}$O nuclei as consisting from different subsystems. First, such a
cluster configuration allows one to consider these nuclei as a core
$^{12}$C comprised of three alpha-particles and two valence
nucleons (neutrons or protons). Thus our model accounts for a
more flexible description of the $^{12}$C inside the nuclei
$^{14}$C and $^{14}$O, since the shape of triangle connecting the
centers of mass of three alpha-particles can be varied by valence
neutrons (protons). Second, one can consider the nucleus
$^{14}$C ($^{14}$O) as a ``core'' $^{6}$He($^{6}$Be) surrounded by
two alpha-particles. In this case, triangle connecting center of
mass of an alpha-particle and two valence neutrons can be essentially deformed
by two valence alpha-particles.

In the present paper we are going to study the ground and the first excited $0^{+}$ states of $^{14}$C and $^{14}$O nuclei.
Different types of the cluster-cluster correlations will be calculated and analyzed in order to reveal
the nature of states under investigation.

The nucleus $^{14}$C attracts large attention from both theoretical (see Refs.
\cite{2010PhRvC..82d4301S,
2015PhRvC..91a4315K,
2016EPJA...52..373K,
2016PhRvC..94d4303B,
UJP2016,
2017JPhCS.863a2033B,
2017PhRvC..95f4318B,
UJP2017})
and experimental (see Refs.
\cite{1993NuPhA.564....1T,
2014PhRvC..90e4324F,
2016ChPhC..40k1001T,
2016PhRvC..93d4323B,
2016PhRvC..93a4321F,
2017NCimC..39..372S,
2017PhLB..766...11Y,
2018JPhCS.966a2040D})
points of view. Some of these investigations are devoted to searching of linear configurations
of clusters in these nuclei. The theoretical investigations are mainly
performed within the framework of the antisymmetrized molecular dynamics (AMD).

Preliminary results of the present calculations carried out for the ground state only, are reported in Refs.
\cite{UJP2016}
and \cite{UJP2017}.

\begin{figure}
[ptbh]
\begin{center}
\includegraphics[
width=\textwidth]%
{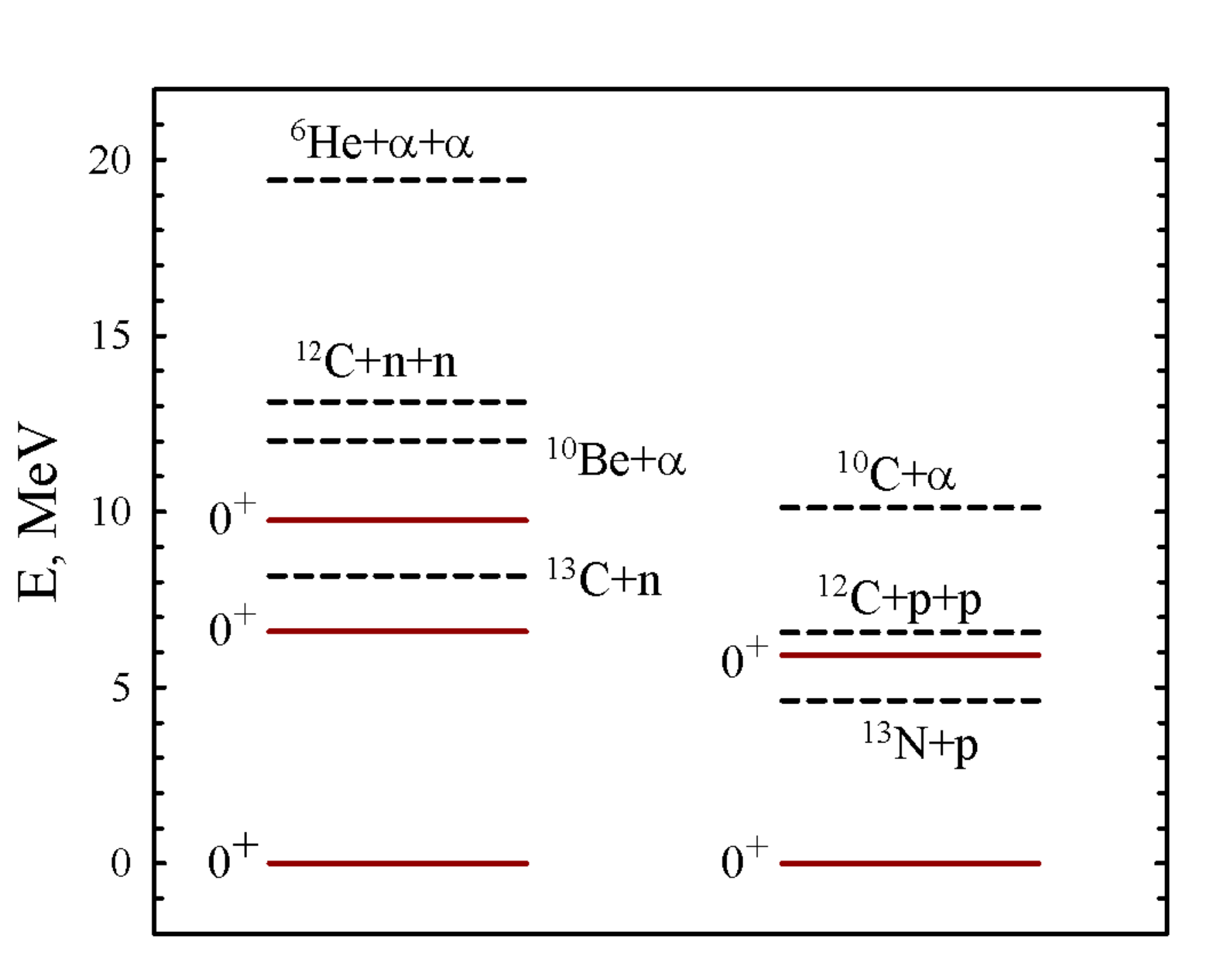}%
 \caption{Spectrum of the lowest $0^{+}$ states of the nuclei $^{14}$C (on the left) and $^{14}$O (on the right), and two- and three-cluster dominant decay channels. Energies in MeV are accounted from the ground states of the nuclei.}%
\label{Fig:1}
\end{center}
\end{figure}

Experimental spectrum of the $0^{+}$ states of $^{14}$C is presented in Table
\ref{Tab:1}
and is taken from Ref. \cite{1993NuPhA.564....1T}.
\begin{table}[tbp] \centering
\caption{Experimental spectrum of the lowest $0^+$ states and energies of two- and three-cluster thresholds in $^{14}$C and $^{14}$O.}%
\vskip3mm
\begin{tabular}
[c]{cccccc}\hline $^{14}$C & $E^{*}\!\left(0^{+}\!\right)\!-\!E_{0}\!\left(
0^{+}\!\right) $, MeV & 6.589 & 9.746 &  &
\\
& $E_{threshold}$, MeV & 8.177 & 12.012 & 13.123 & 19.424
\\
& Decay channel & $^{13}$C$+n$ & $^{10}$Be$+\alpha$ & $^{12}$C$+n\!+\!n$ & $^{6}
$He$+\alpha\!+\!\alpha$  \\\hline
$^{14}$O & $E^{*}\!\left(0^{+}\!\right)\!-\!E_{0}\!\left(
0^{+}\!\right) $, MeV & 5.920 &  &  &
\\
& $E_{threshold}$, MeV & 4.628 & 10.118 & 6.572 &
\\
& Decay channel & $^{13}$N$+p$ & $^{10}$C$+\alpha$ & $^{12}$C$+p\!+\!p$ &  \\\hline
\end{tabular}
\label{Tab:1}
\end{table}%
 In Figure \ref{Fig:1},
we show the spectrum of the $0^{+}$ states in $^{14}$C and the dominant
two- and three-cluster decay thresholds. This figure suggests certain hierarchy
of two- and three-cluster channels in $^{14}$C. The most important
two-body channel is the $^{13}$C$+n$ channel. One may conjecture
that this channel is responsible or, in other words, dominant for
formation of the ground and first excited $0^{+}$ states, and the
$0^{+}$ resonance states which lies just above the threshold of
this channel. Second of importance is the $^{10}$Be$+\alpha$
channel and then the three-body $^{12}$C$\,+\,n+n$ channel. In the
$^{14}$O nucleus, there is only one excited $0^{+}$ state which
lies between the two-body $^{13}$N$+p$ and three-body
$^{12}$C$+p+p$ thresholds.

The layout of the present paper is the following. In Section
\ref{Sec:Method}, we shortly explain the main ideas and key
elements of the five-cluster model used. The key elements include
an alpha-alpha potential, the nucleon-nucleon one in the singlet
state, and alpha-nucleon potential as well. In Section
\ref{Sec:Results}, we formulate the method of finding the wave
functions for the ground and excited states of $^{14}$C and
$^{14}$O nuclei within the five-cluster approach. A detailed analysis of the results obtained is also
presented in this section.

\section{Five-cluster model for $^{14}$O and $^{14}$C nuclei \label{Sec:Method}}

In this section, we formulate our five-cluster model and describe
the method of calculations.

For $^{14}$O nucleus considered as three $\alpha$-particles and
two extra protons, we start from a model Hamiltonian \cite{UJP2016}:
\begin{eqnarray}
\hat{H}  &= & \sum\limits_{i=1}^{2} \frac{\mathbf{p}_{i}^{2}}{2m_{p}}+ \sum
\limits_{i=3}^{5} \frac{\mathbf{p}_{i}^{2}}{2m_{\alpha}}+U_{pp} \left(
r_{12}\right) +
   \sum\limits_{j>i=3}^{5}  \hat{U}_{\alpha\alpha} \left(
r_{ij}\right) \nonumber \\
&+& \sum\limits_{i=1}^{2} \sum\limits_{j=3}^{5}\hat{U}_{p\alpha}  \left(
r_{ij}\right)
+   \sum\limits_{j>i=1}^{5}  \frac{Z_{i}Z_{j}e^{2}}{r_{ij}}.
\label{E1}%
\end{eqnarray}
Within our approach, we make some approximations to simplify the model. In particular, considering the $0^{+}$ states, we take into account only central part of interactions between particles, and this means that the total spin of the nucleus is a good quantum number.
We restrict ourselves with the zero value of the total spin, which in our model coincides with the spin of two valence nucleons. It is then naturally to involve nucleon-nucleon interaction in the singlet state only. An account of an interaction of two nucleons in the triplet state may be important for highly excited states of the nuclei under consideration. We also neglect the spin-orbit part of the $\alpha–N$ interaction assuming that it plays minor role in formation of the $0^{+}$ states in $^{14}$C and $^{14}$O nuclei.
Starting with the Hamiltonian (\ref{E1}) for “point-like” particles, we nevertheless assume that the constituent particles have a finite size, and that they have definite charge and density distributions known from the experiment, and we take them into account within the Helm approximation \cite{1956PhRv..104.1466H} after the solution of the five-body problem (\ref{E1}) to study the charge and mass distributions, as well as the charge form factors of the nuclei.
Note also that the Coulomb repulsion in (\ref{E1}) could be chosen in a modified form at short distances with an account of finite size of particles. But, as it follows from comparison of mirror nuclei $^{10}$Be and $^{10}$C \cite{2014PAN....77..415G,2011NPAE}, as well as $^{14}$C and $^{14}$O \cite{UJP2016}, the Coulomb interaction between particles does not play the decisive role in the structure of the nuclei as compared to the nuclear forces. Moreover, due to the phase volume $\sim r^{2}dr$  in the integrals, a contribution of the Coulomb repulsion at short distances into the matrix elements is small. Thus, our choice of simple form of Coulomb repulsion at short distances does not play essential role. More important is the fact that, in our simple model, we ignore the relativistic effects in the kinetic energy operator and tensor nuclear forces. We also do not take into account other possible cluster configurations in the nuclei essentially important for highly excited states. The role of all these effects needs a separate consideration and analysis.

One can explicitly extract the kinetic energy of the center of
mass of the nucleus from expression (\ref{E1}), but there is no
sense to carry out this trivial procedure since we use the
translation invariant wave functions (see below) giving
automatically the zero kinetic energy of the center of mass of the
system. The Hamiltonian for $^{14}$C nucleus within the
five-cluster model (three $\alpha$-particles plus two extra
neutrons) is very similar to (\ref{E1}), but with neutron mass
$m_{n}$ instead of the proton one in the kinetic energy, and less
strong Coulomb repulsion (due to the fact that $Z_{1}=Z_{2}=0$ for
extra neutrons in $^{14}$C nucleus, while $Z_{1}=Z_{2}=1$ for
extra protons in $^{14}$O, and in both cases $Z_{3}=Z_{4}=Z_{5}=2$
\,for $\alpha$-particles). We assume the nuclear potential between the extra
protons $U_{pp}$ (in $^{14}$O) to be the same as the one between
extra neutrons $U_{nn}$ (in $^{14}$C). This assumption is in concordance with
the charge independence of nuclear forces. This interaction potential
in the singlet state proposed in \cite{2009PAN....72....6G} is chosen in the form
\begin{equation}
U_{nn}\left(r\right)=U_{pp}\left(r\right)=\sum\limits_{k=1}^{3}V_{k}\cdot\exp\left(-\left(\frac{r}{R_{k}}\right)^{2}\right),
\label{E2}%
\end{equation}
where $V_{1}=952.15$ MeV, $R_{1}=0.44$ fm; $V_{2}=-79.39$ MeV,
$R_{2}=0.959$ fm; and $V_{3}=-37.89$ MeV, $R_{3}=1.657$ fm. The
potential (\ref{E2}) leads to a description of the low-energy neutron-neutron phase
shift in the singlet state with reasonable accuracy, including the
low-energy scattering parameters. It was successfully
used for extra nucleon interaction in studying $^{6}$He
\cite{2009PAN....72....6G}, $^{10}$Be and $^{10}$C \cite{2014PAN....77..415G, 2011NPAE} nuclei. The
same potential was used for a study of the ground state structure
of $^{14}$C and $^{14}$O nuclei \cite{UJP2016}.

As to the interaction potential between an extra neutron and an
$\alpha$-particle, we use the generalized type potential
with local (attraction) and non-local
(separable repulsion) terms. This type of potentials \cite{1971PhLB...34..581N,ElChYa1979} enables one
to concord the experimental phase shifts (having difference in
$\pi$ between their values at zero and infinite energies) and
the fact that neutron and $\alpha$-particle have no bound states.
This is possible due to a generalized Levenson's theorem for such
type of potentials \cite{ElChYa1979}. Parameters of this potential are
given in \cite{UJP2016}. The $p\alpha$-interaction
potential is almost the same as the $n\alpha$ one, but with a
little bit modified parameters \cite{UJP2016}. The
small difference in parameters can be justified by the fact that,
within an $\alpha$-particle, there is a little bit different (of
greater radius) distribution of protons as compared to that of
neutrons, mainly due to the Coulomb interaction \cite{UJP2007}.

For $\alpha\alpha$-interaction, we also use the generalized
potentials with a local part containing repulsion and attraction,
and with additional non-local (separable) repulsion. Parameters of
the potential are given in \cite{UJP2016}. The total set of potentials
used in Hamiltonian (\ref{E1}) is selected in such a way that to
reproduce experimental values (binding energy and charge radius)
of the ground state of $^{14}$C nucleus simultaneously with the
corresponding phase shifts (at least with qualitative accuracy)
for corresponding pair collisions of particles constituting the
nucleus within the five-particle model. As to $^{14}$O nucleus,
its experimental charge radius is unknown (see our theoretical
estimation \cite{UJP2016}), thus we adjusted the potentials to
reproduce only the binding energy of $^{14}$O.

The structure functions of the ground state of the both $^{14}$C
and $^{14}$O nuclei within the five-particle model were studied in
detail in \cite{UJP2016}. Now we are going to study the first
excited 0$^{+}$ state of these nuclei within the same variational
method with Gaussian basis \cite{1977JPhG....3..795K, SuzukiVarga,
2003PrPNP..51..223H}. A study of higher excited
states needs some other approaches since they are higher than the
first $^{13}$C$+n$ threshold. Moreover, treating the nuclei as
consisting of $\alpha$-particles ignores a contribution of their
excitation or breakup processes, and thus it has sense only for
the lowest states of $^{14}$C and $^{14}$O nuclei to be
considered within our model.

The spectrum and wave functions of the system are found with high
accuracy within the variational method with the use of Gaussian
basis \cite{1977JPhG....3..795K,SuzukiVarga, 2003PrPNP..51..223H,
2018FrPhy..13.2106H}. For $J^{\pi}=0^{+}$ states, the coordinate part
of the wave functions of the five-particle system can be found in the form
\begin{equation}
\Phi=\hat{S}\sum_{k=1}^{K}C_{k}\varphi_{k}\equiv\hat{S}\sum_{k=1}^{K}C_{k}\,\exp \left( -\sum_{j>i=1}^{5}
a_{k,ij}\left(\mathbf{r}_{i}-\mathbf{r}_{j}\right)^{2} \right)  ,\label{E3}%
\end{equation}
where $\hat{S}$ is the symmetrization operator with respect to
identical $\alpha$-particles, as well as to identical extra nucleons (since, according to our assumption, they are in the singlet state, and their spin function and thus the total function are antisymmetric with respect to the extra nucleons permutation). The spin part of the total wave function is not indicated explicitly for brevity. In all the calculations of matrix elements it gives simply a unitary multiplier. In (\ref{E3}), $K$ is a dimension of the basis. The
coefficients $C_{k}$ can be found within the Galerkin method from
the algebraic system of equations
\begin{equation}
\sum_{m=1}^{K}C_{m}\left\langle  \hat{S}\varphi_{k}\left\vert
\hat {H}-E\right\vert \hat{S}\varphi_{m} \right\rangle =0,\quad
k\,=\,0,1,...,K.\label{E4}%
\end{equation}
Matrix elements for the Hamiltonian (\ref{E1}) are known
to have an explicit form due to the Gaussian functions of the
basis. Within the complete basic set, at $K\rightarrow\infty$, the
system (\ref{E4}) is equivalent to the Schr\"{o}dinger equation.
An advantage of the Gaussian bases lies in the fact that, at a
definite $K$, one can achieve better accuracy by varying the
non-linear parameters $a_{k,ij}$. This needs more time for
calculations, but results in the best accuracy for the energies
and wave functions at a given $K$. The most efficient method to
save the time of calculations is the random choice of the parameters $a_{k,ij}$
and the increasing of $K$ (``variational method with stochastic
Gaussian bases'' \cite{1977JPhG....3..795K,SuzukiVarga}). We used the both approaches. As a
result, we have energies and the wave functions of the ground and
first excited $0^{+}$ states of $^{14}$C and $^{14}$O nuclei
within the five-particle model with sufficiently high accuracy.
After calculation of r.m.s. radii and density distributions for “point-like” particles, we calculate the charge r.m.s. radii and charge density distributions taking into account \cite{UJP2016,UJP2017} the size of $\alpha$-particles and that of extra nucleons within the Helm approximation \cite{1956PhRv..104.1466H}. Only after this correction, we compare the calculated charge radii with the experimental data. The same Helm correction is used for calculation of both elastic and transition electric form factors (see below).

\section{Results and Discussion \label{Sec:Results}}

In this section we present and discuss main results of our
investigations.

In Table \ref{Tab:2},
results for energies and charge r.m.s. radii are given for both $^{14}$C and $^{14}$O
nuclei calculated for the ground and first excited $0^{+}$ states.
The charge radii are found within the Helm approximation.
Experimental values of the ground state energies of the both
nuclei, and charge r.m.s. radius of $^{14}$C nucleus, are
reproduced with high accuracy \cite{UJP2016} by adjusting the
parameters of potentials. The result $2.415$ fm for the charge
radius of $^{14}$O nucleus is our prediction \cite{UJP2016}.

The first excited $0^{+}$ state energy of $^{14}$O nucleus is
calculated with $K=500$ functions of the Gaussian basis, and it is
seen to be in a good coincidence with the experiment. The similar
result for $^{14}$C nucleus is obtained with $K=600$ Gaussian
functions. It is in a qualitative agreement with the experimental
energy of the first $0^{+}$ excited state. We see that the set of
potentials proposed in \cite{UJP2016} used for $^{14}$O nucleus
appeared to be more happy choice to treat the excited $0^{+}$
state than that used for $^{14}$C one. The results for charge
r.m.s. radii of the excited state are seen to be a little bit
greater than the ones for the ground state. Note also, that for
the ground state we have $R_{ch,0}(^{14}$O$)<R_{ch,0}(^{14}$C$)$,
since the extra protons of $^{14}$O nucleus move mainly inside the
$^{12}$C cluster \cite{UJP2016}. For the considered excited states,
one has more natural situation $R_{ch,1}(^{14}$O$)>R_{ch,1}(^{14}$C$)$.

\begin{table}[tbp] \centering
\caption{Experimental (E) and calculated (T) energies of the ground and the first
excited $0^{+}$ states and corresponding
charge radii for $^{14}$C and $^{14}$O nuclei.}%
\vskip3mm
\begin{tabular}
[c]{ccccc}\hline  & $^{14}$C (T)
& $^{14}$C (E) &
$^{14}$O (T) & $^{14}$O (E)
\\\hline
$E_{0}$, MeV & -20.398 & -20.3977 &
-13.845 & -13.8445
\\
$R_{ch}$, fm & 2.500 & 2.496 \cite{1982NuPhA.379..523S}, 2.503 \cite{2013ADNDT..99...69A} &
2.415 & --
\\
$E_{1}$, MeV & -13.223 & -13.808 & -7.932
& -7.925
\\
$R_{ch}$, fm & 2.582 & -- & 2.589 & --
\\\hline
\end{tabular}
\label{Tab:2}
\end{table}%

More information about the spatial structure of the excited states
under consideration can be obtained from the analysis of r.m.s.
relative distances between constituting particles of the nuclei.
These distances ($r_{\alpha\alpha}$ between $\alpha$-particles,
$r_{NN}$ -- between extra nucleons, and $r_{N\alpha}$ -- between
an extra nucleon and an $\alpha$-particle) are presented in Table
\ref{Tab:3}.
For comparison, the results \cite{UJP2016} for the
ground state are given too. The general increase of the
relative distances between particles for the excited state is
observed as compared to the ones for the ground state. This fact is in
concordance with the fact that, generally speaking, the less is
the binding energy, the greater are the distances between
particles. The greater relative distances mean better
substantiation of the five-particle model for the excited states
considered, since particles of finite size are less overlapping in
these states then in the ground state.

In the same Table \ref{Tab:3},
we present also the r.m.s. radii (the r.m.s. distances from the center of mass of a
nucleus) calculated for ``point-like'' $\alpha$-particles
($R_{\alpha}$) and extra nucleons ($R_{N}$). These values
calculated directly are known to be connected with r.m.s. relative
distances by the identity \cite{UJP2017}:
\begin{equation}\label{E5}
R_{i}^{2} = \frac{1}{M^{2}}\left(\left(M-m_{i}\right)\sum_{\footnotesize{\begin{array}{c} {j\neq i}
\end{array}}}
  m_{j}r_{ij}^{2}   -     \sum_{\footnotesize{\begin{array}{c} {j<k}
\\ {\left(j\neq i, k\neq i\right)}
\end{array}}}         m_{j}m_{k}r_{jk}^{2}  \right),
\end{equation}
where $M=\sum\limits_{j}m_{j}$ is the total mass of the system.

\begin{table}[tbp] \centering
\caption{Relative r.m.s. distances between particles, and r.m.s. radii in
$^{14}$C and $^{14}$O nuclei (in fm), calculated within the five-particle
model for “point-like” particles. The last column presents the mass radii $R_{m}$ with an account of the size of particles within the Helm approximation.}
\vskip3mm
\begin{tabular}
[c]{ccccccc}\hline &
$r_{\alpha\alpha}$       &
      $r_{NN}$       &
      $r_{N\alpha}$      &
      $R_{\alpha}$      &
      $R_{N}$      &
      $R_{m}$
\\\hline
$^{14}$C (ground $0^{+}$) & 3.189 & 2.621 &
2.667 & 1.852 & 1.786 & 2.433
\\
$^{14}$C (1st excited $0^{+}$) & 3.367 & 3.211 &
3.197 & 1.961 & 2.328 & 2.568
\\
$^{14}$O (ground $0^{+}$) & 3.239 & 2.732 & 2.750  & 1.882 &
1.865 & 2.461
\\
$^{14}$O (1st excited $0^{+}$) & 3.412 & 3.305 & 3.265  & 1.991 &
2.387 & 2.596
\\\hline
\end{tabular}
\label{Tab:3}
\end{table}%

\begin{figure}
[hptb] 
\begin{center}
\includegraphics[
width=\textwidth]%
{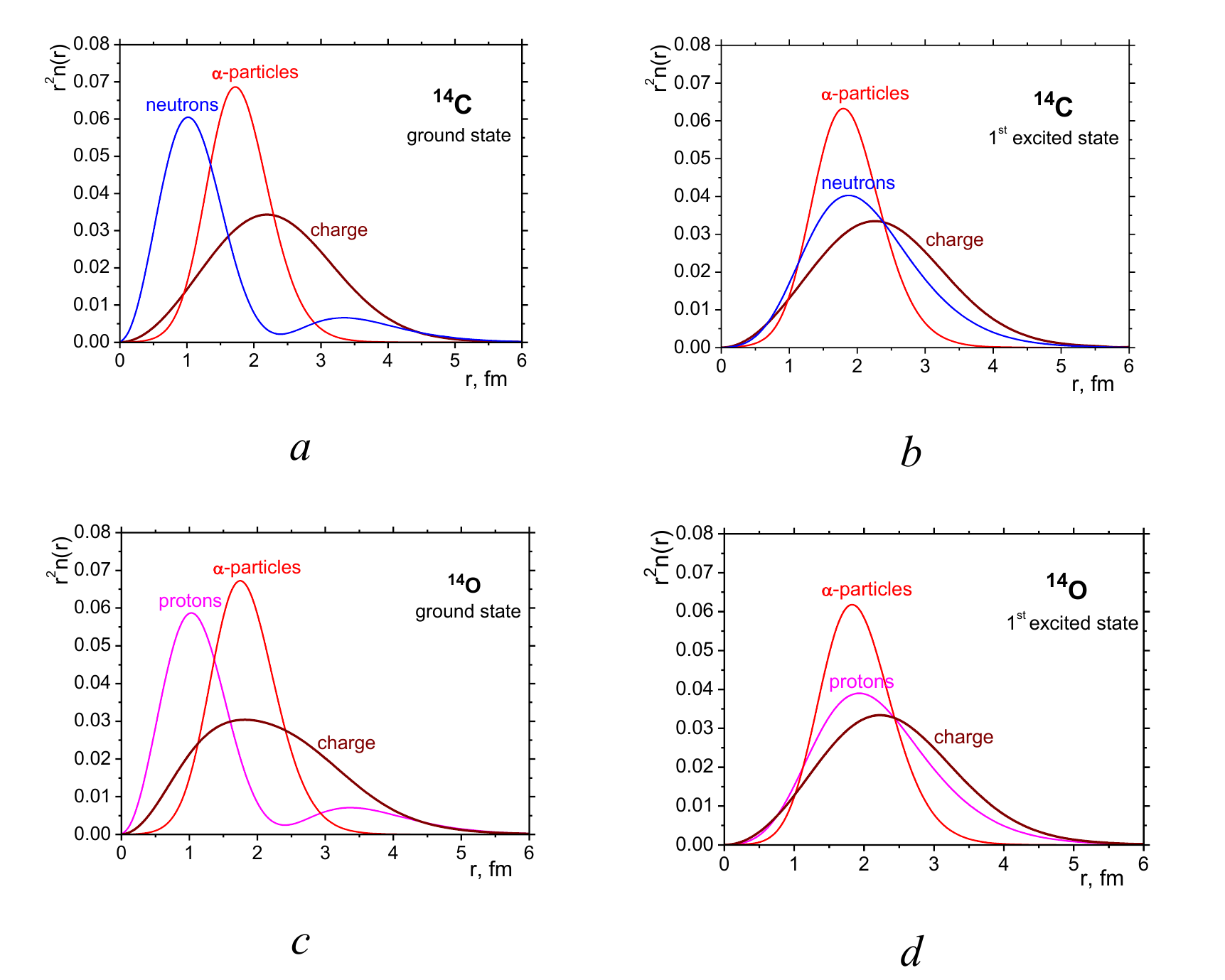}%
\caption{Density distributions (multiplied by $r^{2}$) of $\alpha$-particles and extra nucleons in the ground ($a$, $c$) and first excited ($b$, $d$) \, $0^{+}$  states of $^{14}$C and $^{14}$O nuclei calculated for “point-like” particles. The calculated charge density distribution (with an account of finite size of constituent particles in Helm approximation) is depicted by thick line.}
\label{Fig:2}%
\end{center}
\end{figure}

Now consider the density distributions of particles in the excited
state in comparison with those for the ground state of the nuclei
$^{14}$C and $^{14}$O. In Fig.\ \ref{Fig:2}, density
distributions multiplied by $r^{2}$ are shown. It is clearly seen
that, for the ground state of the both nuclei, the extra nucleons
move mainly inside the $^{12}$C cluster, while it is not so for
the excited state. That is why the charge density distribution
calculated for the ground state of $^{14}$O nucleus has maximum at
shorter distances than the similar distribution for $^{14}$C does
(since there are charged extra protons in $^{14}$O nucleus moving
more close to the center of the nucleus than the $\alpha$-particles do, while in
$^{14}$C nucleus, only $\alpha$-particles make a contribution to the charge distribution). This results in mentioned above inequality
$R_{ch,0}(^{14}$O$)<R_{ch,0}(^{14}$C$)$. And this effect is not
revealed for the excited states.

\begin{figure}
[hptb] 
\begin{center}
\includegraphics[
width=\textwidth]%
{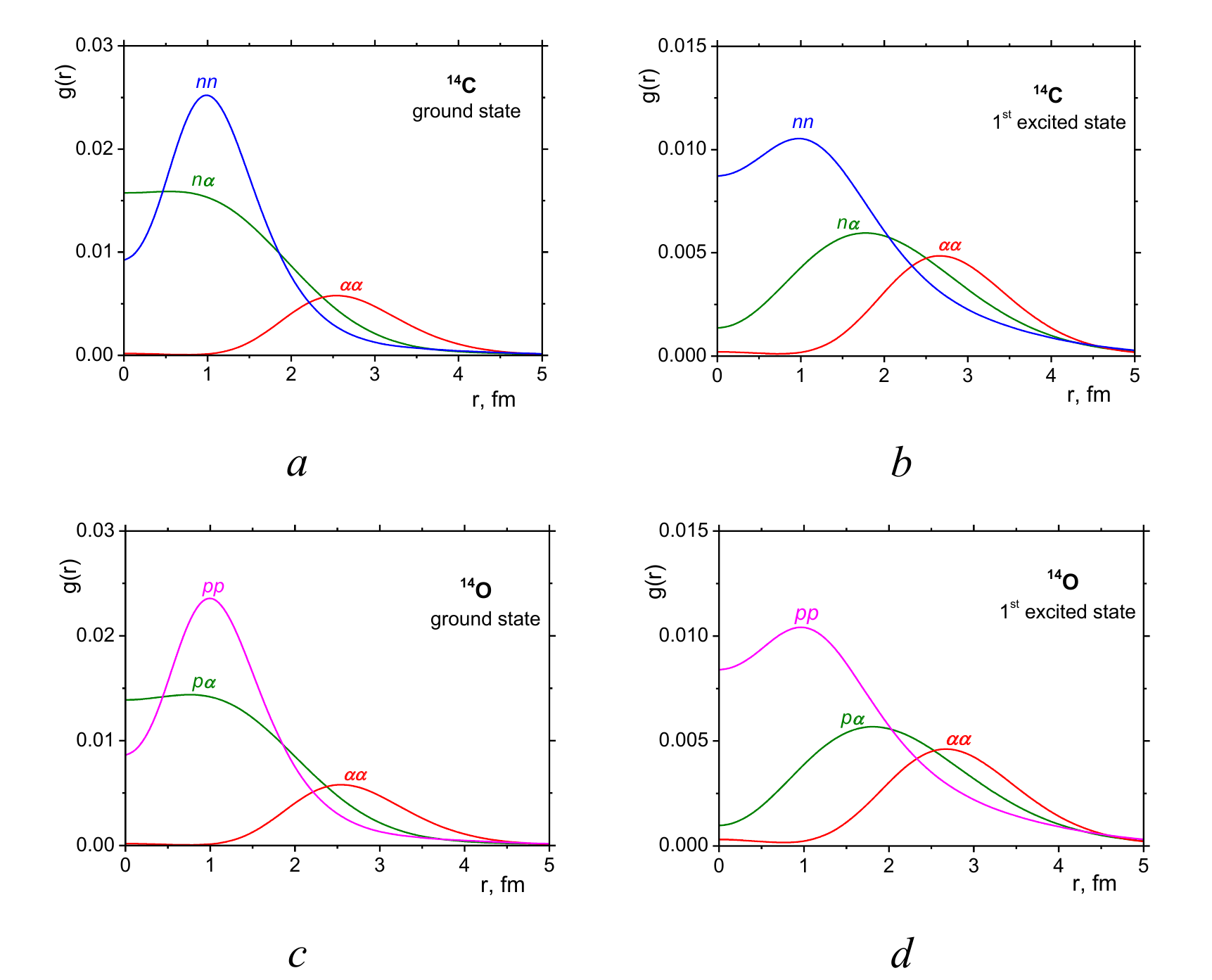}%
 \caption{Pair correlation functions for the ground ($a$, $c$) and
first excited ($b$, $d$)    $0^{+}$ states of $^{14}$C and $^{14}$O nuclei.}
\label{Fig:3}  
\end{center}
\end{figure}

In Fig.\ref{Fig:3},
we depict the pair correlation functions for the same nuclei in
the ground and the first excited states. The obtained distributions
show that, in the excited state, the average distances of extra
nucleons from $\alpha$-particles are enlarged as compared to those
in the ground state. This fact is in complete concordance with the results
obtained for r.m.s. distances (see Table \ref{Tab:3}).

\begin{figure}
[hptb] 
\begin{center}
\includegraphics[
width=\textwidth]%
{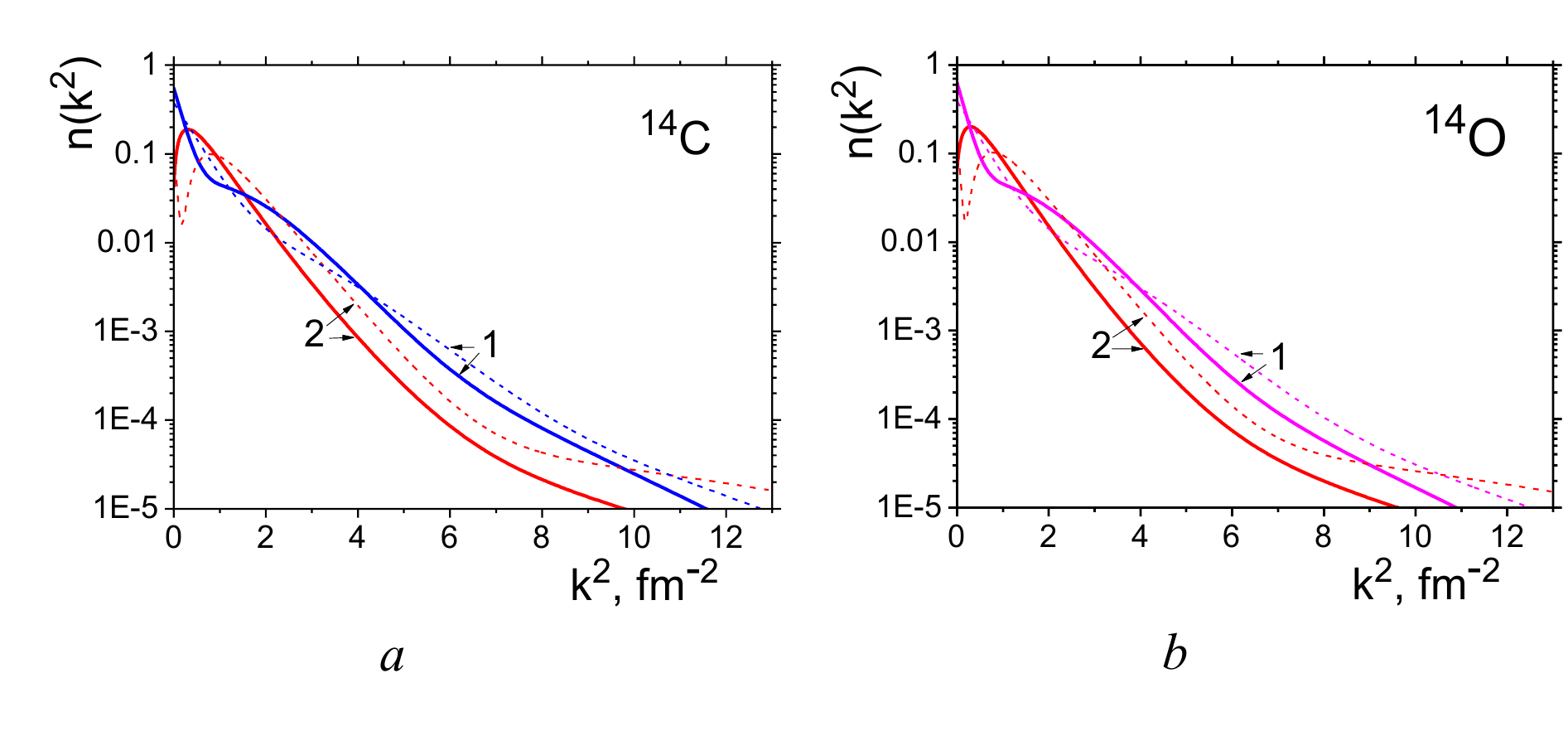}%
\caption{Momentum distributions in the ground (dashed lines) and
first excited (solid lines)   $0^{+}$ states of $^{14}$C and $^{14}$O nuclei
($a$ and $b$, respectively). Curves $1$ -- for $\alpha$-particles,
and curves $2$ -- for extra nucleons.}
\label{Fig:4} 
\end{center}
\end{figure}

Consider the momentum distributions of extra nucleons and
$\alpha$-particles in the ground and first excited $0^{+}$ states of the
both nuclei. In Fig. \ref{Fig:4}, we compare the momentum distributions of the $^{14}$C and $^{14}$O
nuclei. The dashed lines depict the momentum distributions for the
ground state of these nuclei \cite{UJP2016}. It is clearly seen for the excited state momentum distributions to have a somewhat
greater low-energy (or low-momenta) values than those in the ground state. This is natural for the momentum distributions of particles of less bound systems as compared to more strongly bound ones. We also note that the momentum distributions of extra nucleons in the excited state have dips shifted towards the zero momenta as compared to the ground state. And the change of regimes of decreasing of the momentum distribution profiles for $\alpha$-particles is more clearly seen for the excited state than for the ground state, and it is observed at lower momenta.

Generally speaking, the mirror nuclei $^{14}$C and $^{14}$O are
seen to have very similar spatial structure functions and the momentum distributions (at least from qualitative
point of view). That is why, for brevity, we consider below some
details of the calculated probability density distributions only
for $^{14}$C nucleus.

\begin{figure}
[hptb] 
\begin{center}
\includegraphics[
width=\textwidth]%
{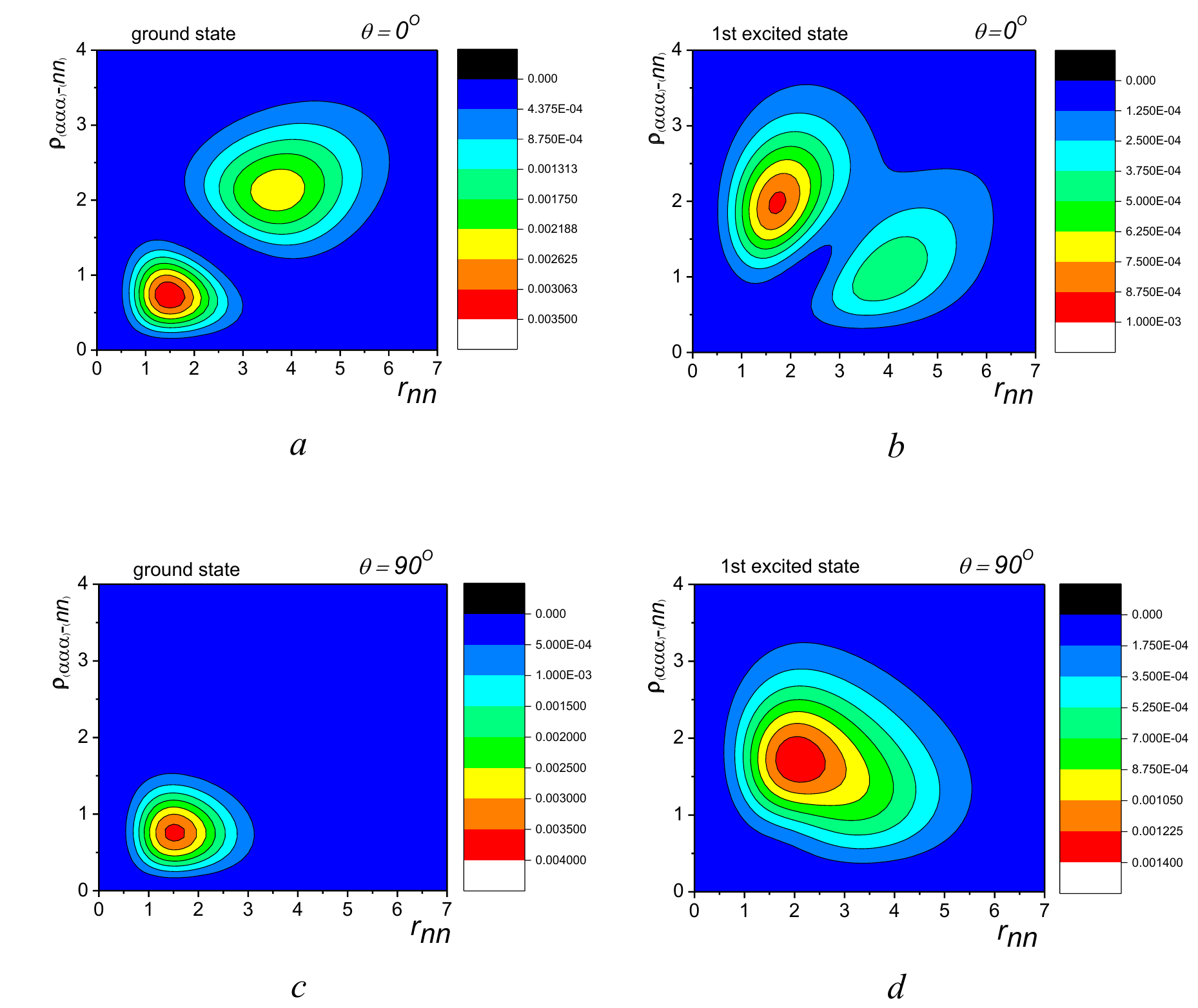}%
 \caption{Probability density distribution $P_{NN}\left(r,\rho,\theta\right)$ for the ground ($a$, $c$) and
first excited ($b$, $d$) $0^{+}$ states of $^{14}$C nucleus, at fixed $\theta=0^{\circ}$ ($a$,$b$) and $\theta=90^{\circ}$ ($c$,$d$).}
\label{Fig:5}%
\end{center}
\end{figure}

It is impossible to depict directly the squared wave function depending on too large number of
variables which is proper to the five-particle system. That is why we introduce the values being the
squared wave function partially integrated over some of the variables and thus depending only on
the rest few of them. In particular, consider the following probability density distribution
\begin{equation}\label{E6}
P_{NN}\left(r,\rho,\theta\right)\equiv r^{2}\rho^{2}
\left\langle \Phi\left\vert
\delta\left(\mathbf{r}-\mathbf{r}_{NN}\right)\delta\left(\mathbf{\rho}-\mathbf{\rho}_{\left(NN\right)\left(3\alpha\right)}\right)
\right\vert \Phi\right\rangle,
\end{equation}
where $r$ is the distance between the extra nucleons, $\rho$ is the distance between the centers of mass of the two-nucleon subsystem and the three-alpha ($^{12}$C) cluster, and $\theta$ is an angle between the vectors $\mathbf{r}$ and $\mathbf{\rho}$. In Fig. \ref{Fig:5}, we show this value at fixed $\theta=0^{\circ}$ and $\theta=90^{\circ}$ in case of $^{14}$C nucleus. Very similar pictures can be drawn for $^{14}$O nucleus since an additional Coulomb repulsion between extra protons themselves and that between the protons and $\alpha$-particles play minor role as compared to the nuclear forces. It is obvious that two peaks \cite{UJP2017} observed for the probability density (\ref{E6}) at $\theta=0^{\circ}$ in the ground $0^{+}$ state change their positions in the first $0^{+}$ excited state. In particular, the dineutron cluster starts to move around the $^{12}$C cluster at greater distances of the order of $2$ fm, while a configuration corresponding to two separate neutrons, vice versa, approaches to the center of mass of the $^{12}$C cluster. At $\theta=90^{\circ}$, the second peak does not reveal itself, but it is seen once more that the two neutrons move further from the center of mass of the $^{12}$C cluster, and it is seen that the size of the two extra neutrons cluster in the excited state becomes greater.

\begin{figure}
[hptb] 
\begin{center}
\includegraphics[
width=\textwidth]%
{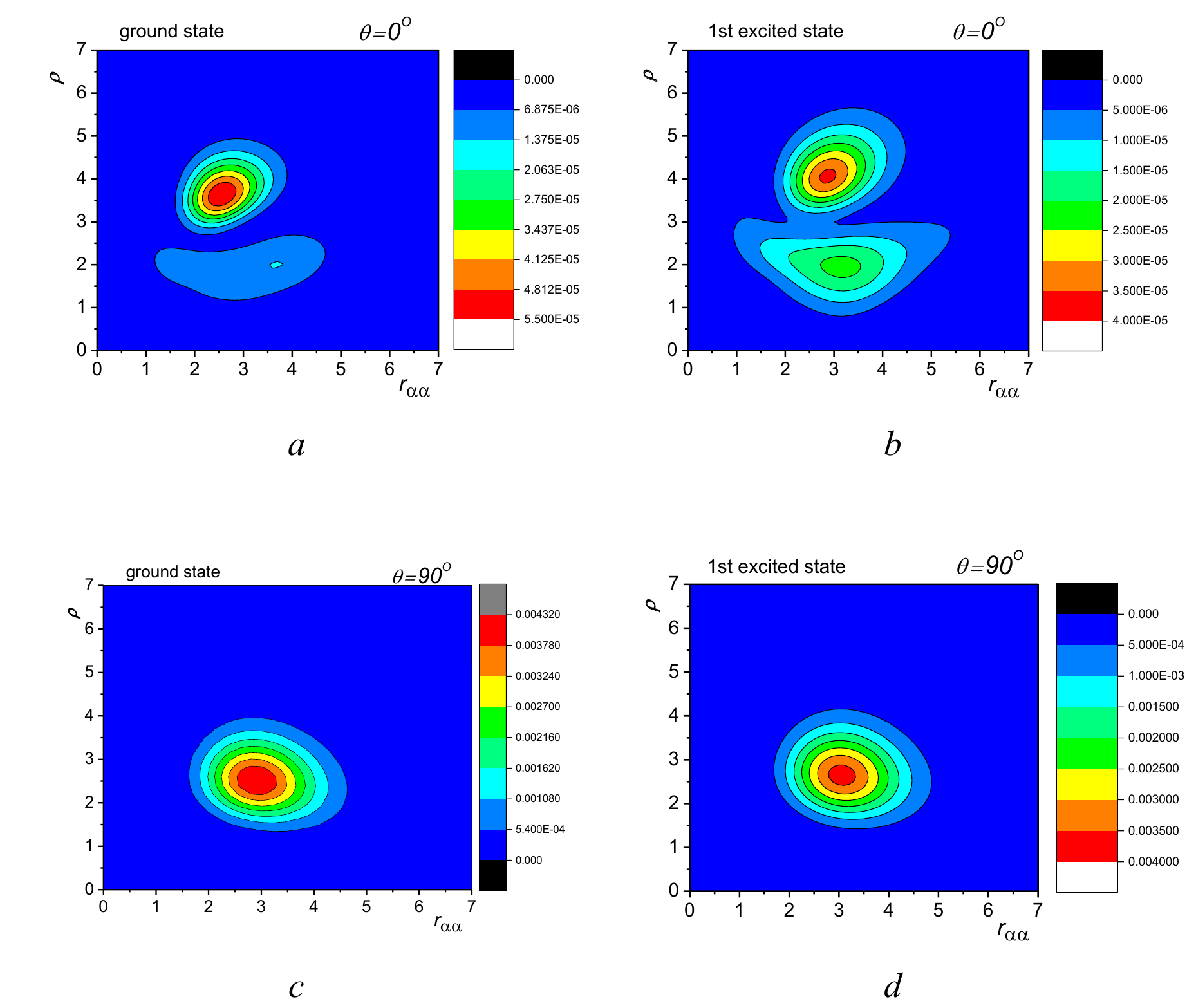}%
 \caption{Probability density distribution $P_{\alpha\alpha}\left(r,\rho,\theta\right)$ for the ground ($a$, $c$) and
first excited ($b$, $d$) $0^{+}$ states of $^{14}$C nucleus.}
\label{Fig:6}%
\end{center}
\end{figure}

We may consider another probability density,
\begin{equation}\label{E7}
P_{\alpha\alpha}\left(r,\rho,\theta\right)\equiv r^{2}\rho^{2}
\left\langle \Phi\left\vert
\delta\left(\mathbf{r}-\mathbf{r}_{\alpha\alpha}\right)\delta\left(\mathbf{\rho}-\mathbf{\rho}_{\left(\alpha\alpha\right)\left(^{6}He\right)}\right)
\right\vert \Phi\right\rangle,
\end{equation}
where $r$ is now a distance between two $\alpha$-particles, and $\rho$ is the distance between the center of mass of the two $\alpha$-particle cluster and the center of mass of the rest particles (of the $^{6}$He cluster). In Fig. \ref{Fig:6}, we show this value for the same nucleus in the ground $0^{+}$ and the first $0^{+}$ excited states. It is obvious from the pictures that the configurations described by the value (\ref{E7}) almost do not change their positions and form, on the contrary to the previous value (\ref{E6}) depicted in Fig.\ref{Fig:5}. This means that the $^{12}$C cluster is practically not excited in the $0^{+}$ first excited state, while the excitation is connected with the two-nucleon subsystem of $^{14}$C nucleus. We believe that first this effect was demonstrated for $^{14}$C nucleus in \cite{2010PhRvC..82d4301S} within an approach with the account of all nucleon degrees of freedom.

We also present here the transition electric form factors of $^{14}$O and $^{14}$C nuclei. The elastic form factors of these nuclei and their charge distributions in the ground state are discussed in detail in \cite{UJP2016}. Now we calculate the electric transition form factors for both of these nuclei using the wave functions of the ground $\Phi_{0}$ and the first excited $\Phi_{1}$ states found above.

We remind that, for a ``point-like'' charged particle $k$ of the system described by the wave function $\Phi_{0}$, a contribution to the elastic form factor $F_{k,00}\left(q\right)$ is known to be
\begin{equation}\label{E8}
F_{k,00}\left(q\right)\equiv \int e^{-i\left(\mathbf{q}\mathbf{r}\right)}n_{k}\left(r\right)d\mathbf{r}=
\left\langle\Phi_{0}\left|e^{-i\left(\mathbf{q}\left(\mathbf{r}_{k}-\mathbf{R}_{c.m.}\right)\right)}\right|\Phi_{0}\right\rangle,
\end{equation}
where $n_{k}\left(r\right)$ is the probability density distribution of the particle $k$. For the transition electric form factor between the states $\Phi_{0}$ and $\Phi_{1}$, the same charged ``point-like'' particle makes a contribution $F_{k,01}\left(q\right)$ calculated in a similar way:
\begin{equation}\label{E9}
F_{k,01}\left(q\right)=
\left\langle\Phi_{0}\left|e^{-i\left(\mathbf{q}\left(\mathbf{r}_{k}-\mathbf{R}_{c.m.}\right)\right)}\right|\Phi_{1}\right\rangle.
\end{equation}
An account of the contributions of all the charged particles results in a sum over all such particles with weights proportional to their charges (the total sum of weights being unity). In addition, one has to take into account that the particles of the system are not ``point-like'', and they have their own form factors. Within the Helm approximation \cite{1956PhRv..104.1466H}, a corresponding contribution of a given particle into the total form factor should be multiplied by the own charge form factor of these particle. In particular, for the transition electric form factor of $^{14}$C nucleus we have (in the Helm approximation, neglecting the contribution of extra neutrons)
\begin{equation}\label{E10}
F_{^{14} \text{C},01}\left(q\right)=F_{\alpha,01}\left(q\right)\cdot F_{\text{ch},^{4}\text{He}}\left(q\right),
\end{equation}
where $F_{\text{ch},^{4}\text{He}}\left(q\right)$ is the experimental form factor of $^{4}$He nucleus \cite{1967PhRv..160..874F}.
For the transition form factor of the $^{14}$O nucleus, one has two terms originating from $\alpha$-particles and extra protons contributions:
\begin{equation}\label{E11}
F_{^{14} \text{O},01}\left(q\right)=\frac{3}{4}F_{\alpha,01}\left(q\right)\cdot F_{\text{ch},^{4}\text{He}}\left(q\right)+\frac{1}{4}F_{p,01}\left(q\right)\cdot F_{\text{ch},p}\left(q\right),
\end{equation}
where $F_{\text{ch},p}\left(q\right)$ is the experimental charge form factor of the proton \cite{1992PhRvL..68.3841B}.

\begin{figure}
[hptb]
\begin{center}
\includegraphics[
width=\textwidth]%
{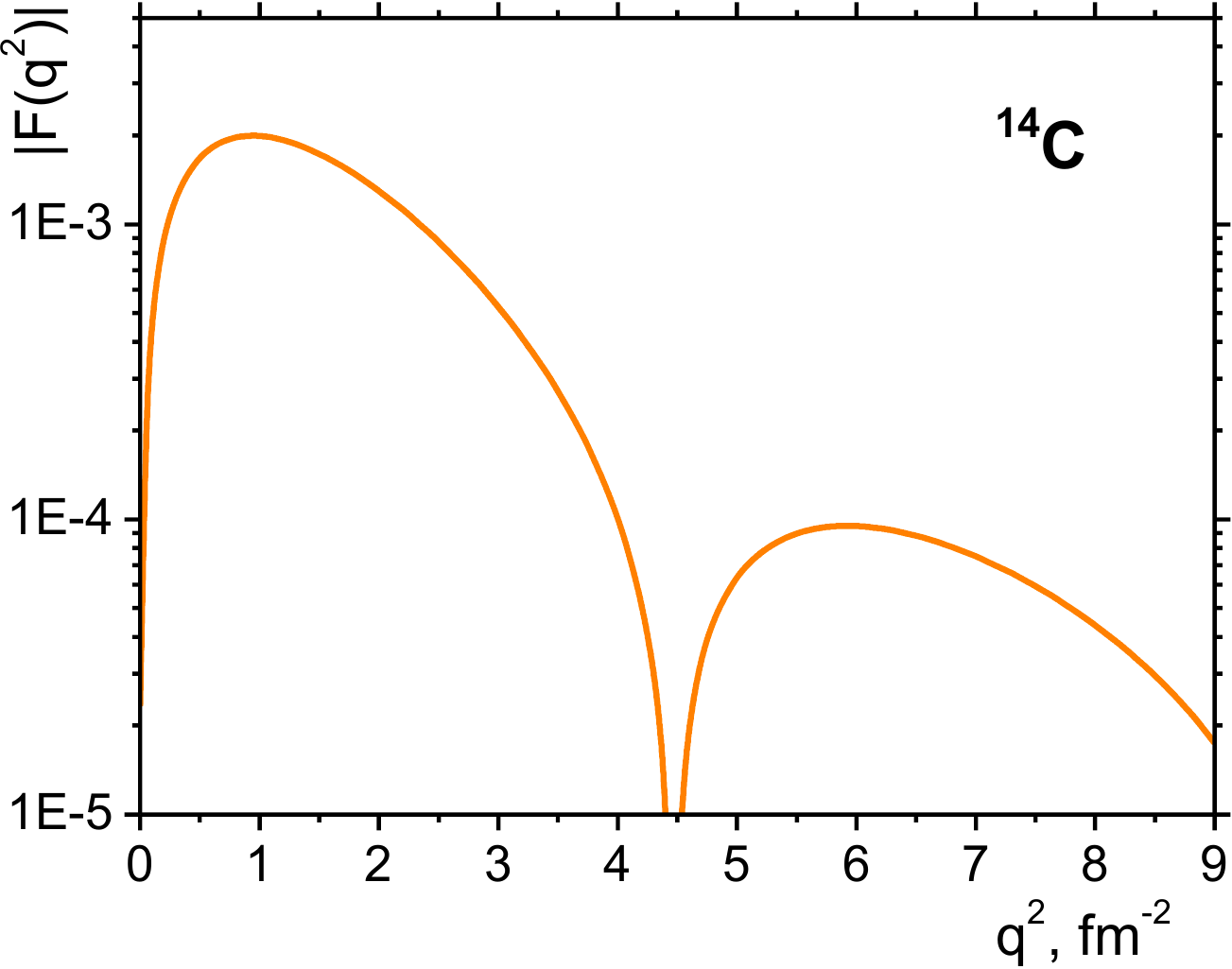}%
 \caption{Transition electric form factor for $^{14}$C nucleus.}
\label{Fig:7}%
\end{center}
\end{figure}

\begin{figure}
[hptb]
\begin{center}
\includegraphics[
width=\textwidth]%
{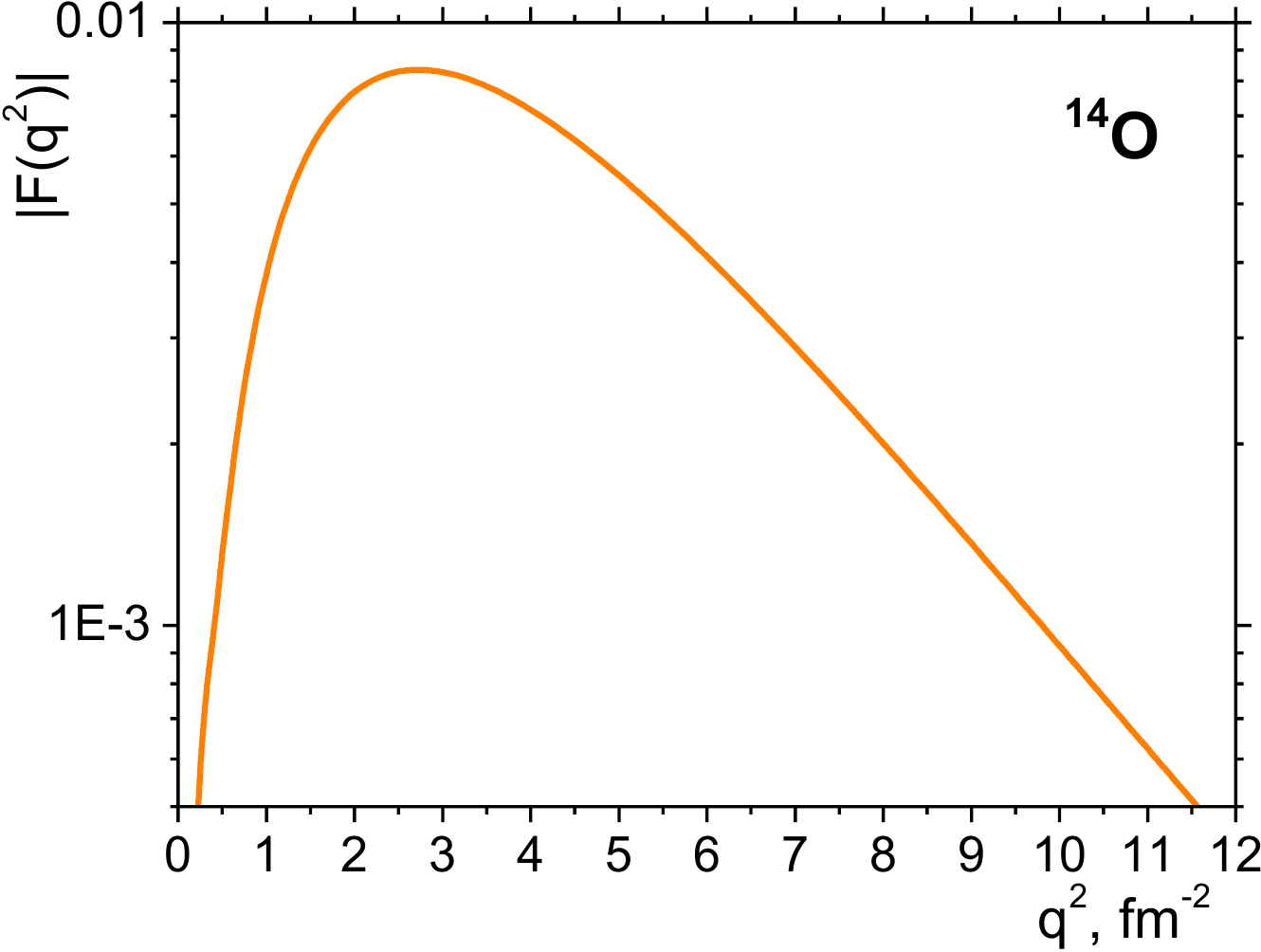}%
 \caption{Transition electric form factor for $^{14}$O nucleus.}
\label{Fig:8}%
\end{center}
\end{figure}

An advantage of the Helm approximation lies in its simple and clear formulation. But, unfortunately, one should take into account, that at large momentum transfer corresponding to short distances in the coordinate space the Helm approximation becomes not accurate enough, because it does not take into account the Pauli principle and exchange effects between overlapping particles. Thus our results for the form factors presented below (as well as the results for elastic form factors obtained earlier \cite{UJP2016}) are reliable at not very high momentum transfer. In Fig. \ref{Fig:7}, the transition electric form factor of $^{14}$C nucleus is presented calculated within approximation (\ref{E10}), while in Fig. \ref{Fig:8} we give the result for $^{14}$O obtained within (\ref{E11}). The dip observed in Fig. \ref{Fig:7} is situated not far from the dip \cite{UJP2016} inherent to the elastic form factor of $^{14}$C nucleus. The presence of such a dip is explained by the property of the first multiplier in the formula (\ref{E10}) for the form factor of $^{14}$C nucleus. On the other hand, the presence of the second term in the expression (\ref{E11}) for the form factor of $^{14}$O nucleus ``smooths out'' the dip proper to the first term. The same was observed for the elastic form factors \cite{UJP2016}.

\section{Conclusions \label{Sec:Conclusions}}
To conclude, we note that the main structure functions (density distributions, form factors, pair correlation functions, and momentum distributions) are studied for the mirror nuclei $^{14}$C and  $^{14}$O within a five-cluster model
(three alpha-particles and two extra nucleons). All the structure functions for the ground and the excited states are compared between themselves, and a conclusion is drawn that the first excited $0^{+}$ state of the both considered nuclei looks like an excitation of the two-nucleon subsystem moving in the field of the $^{12}$C cluster. The both nuclei are shown to have very similar spatial structure, an additional Coulomb repulsion in $^{14}$O playing secondary role. But the electric form factors of these nuclei, due to extra protons present in $^{14}$O, are essentially different (this is true both for the elastic and transition form factors).

\section*{Acknowledgement}
One of the authors (B.E.G.) would like to express appreciation to the European Centre for Theoretical
Studies in Nuclear Physics and Related Areas (ECT*) and the Fondazione Bruno Kessler for financial support and
hospitality within the ECT* ``Ukraine Fellowship Project''.
This work was supported in part by the Program of Fundamental Research of the
Physics and Astronomy Department of the National Academy of Sciences of
Ukraine (Projects No. 0122U000886 and No. 0122U000889).


\begin{thebibliography}{10}
\expandafter\ifx\csname url\endcsname\relax
  \def\url#1{\texttt{#1}}\fi
\expandafter\ifx\csname urlprefix\endcsname\relax\def\urlprefix{URL }\fi
\expandafter\ifx\csname href\endcsname\relax
  \def\href#1#2{#2} \def\path#1{#1}\fi

\bibitem{2010RPPh...73c6301D}
P.~{Descouvemont}, D.~{Baye}, {The R-matrix theory}, Rep. Prog. Phys. 73~(3)
  (2010) 036301.
\newblock \href {http://arxiv.org/abs/1001.0678} {\path{arXiv:1001.0678}},
  \href {https://doi.org/10.1088/0034-4885/73/3/036301}
  {\path{doi:10.1088/0034-4885/73/3/036301}}.

\bibitem{2012PThPS.192....1H}
H.~{Horiuchi}, K.~{Ikeda}, K.~{Kat{\={o}}}, {Recent Developments in Nuclear
  Cluster Physics}, Prog. Theor. Phys. Suppl. 192 (2012) 1--238.
\newblock \href {https://doi.org/10.1143/PTPS.192.1}
  {\path{doi:10.1143/PTPS.192.1}}.

\bibitem{2015PrPNP..82...78F}
Y.~{Funaki}, H.~{Horiuchi}, A.~{Tohsaki}, {Cluster models from RGM to alpha
  condensation and beyond}, Prog. Part. Nucl. Phys. 82 (2015) 78--132.
\newblock \href {https://doi.org/10.1016/j.ppnp.2015.01.001}
  {\path{doi:10.1016/j.ppnp.2015.01.001}}.

\bibitem{2016arXiv160302574B}
C.~{Beck}, {From the stable to the exotic: clustering in light nuclei}, arXiv
  e-prints (2016) arXiv:1603.02574\href {http://arxiv.org/abs/1603.02574}
  {\path{arXiv:1603.02574}}.

\bibitem{2018arXiv181208013B}
C.~{Beck}, {Clusters in light stable and exotic nuclei}, arXiv e-prints (2018)
  arXiv:1812.08013\href {http://arxiv.org/abs/1812.08013}
  {\path{arXiv:1812.08013}}.

\bibitem{2018RvMP...90c5004F}
M.~{Freer}, H.~{Horiuchi}, Y.~{Kanada-En'yo}, D.~{Lee}, U.-G. {Mei{\ss}ner},
  {Microscopic clustering in light nuclei}, Rev. Mod. Phys. 90~(3) (2018)
  035004.
\newblock \href {http://arxiv.org/abs/1705.06192} {\path{arXiv:1705.06192}},
  \href {https://doi.org/10.1103/RevModPhys.90.035004}
  {\path{doi:10.1103/RevModPhys.90.035004}}.

\bibitem{2020FrPhy..1514401Z}
B.~{Zhou}, Y.~{Funaki}, H.~{Horiuchi}, A.~{Tohsaki}, {Nonlocalized clustering
  and evolution of cluster structure in nuclei}, Frontiers of Physics 15~(1)
  (2020) 14401.
\newblock \href {http://arxiv.org/abs/1905.00788} {\path{arXiv:1905.00788}},
  \href {https://doi.org/10.1007/s11467-019-0917-0}
  {\path{doi:10.1007/s11467-019-0917-0}}.

\bibitem{2020PTEP.2020lA101M}
T.~{Myo}, K.~{Kat{\={o}}}, {Complex scaling: Physics of unbound light nuclei
  and perspective}, Prog. Theor. Exp. Phys. 2020~(12) (2020) 12A101.
\newblock \href {http://arxiv.org/abs/2007.12172} {\path{arXiv:2007.12172}},
  \href {https://doi.org/10.1093/ptep/ptaa101}
  {\path{doi:10.1093/ptep/ptaa101}}.

\bibitem{Saito69}
S.~Saito, {Interaction between Clusters and Pauli Principle}, Prog. Theor.
  Phys. {\bf 41}~(3) (1969) 705--722.
\newblock \href {https://doi.org/10.1143/PTP.41.705}
  {\path{doi:10.1143/PTP.41.705}}.

\bibitem{kn:Saito77}
S.~Saito, {Theory of Resonating Group Method and Generator Coordinate Method,
  and Orthogonality Condition Model}, Prog. Theor. Phys. Suppl. {\bf 62} (1977)
  11--89.
\newblock \href {https://doi.org/10.1143/PTPS.62.11}
  {\path{doi:10.1143/PTPS.62.11}}.

\bibitem{2010PhRvC..82d4301S}
T.~{Suhara}, Y.~{Kanada-En'Yo}, {Cluster structures of excited states in
  $^{14}$C}, Phys. Rev. C 82~(4) (2010) 044301.
\newblock \href {http://arxiv.org/abs/1004.4954} {\path{arXiv:1004.4954}},
  \href {https://doi.org/10.1103/PhysRevC.82.044301}
  {\path{doi:10.1103/PhysRevC.82.044301}}.

\bibitem{2015PhRvC..91a4315K}
Y.~{Kanada-En'yo}, {Proton radii of Be, B, and C isotopes}, Phys. Rev. C 91~(1)
  (2015) 014315.
\newblock \href {http://arxiv.org/abs/1411.0765} {\path{arXiv:1411.0765}},
  \href {https://doi.org/10.1103/PhysRevC.91.014315}
  {\path{doi:10.1103/PhysRevC.91.014315}}.

\bibitem{2016EPJA...52..373K}
M.~{Kimura}, T.~{Suhara}, Y.~{Kanada-En'yo}, {Antisymmetrized molecular
  dynamics studies for exotic clustering phenomena in neutron-rich nuclei},
  Eur. Phys. J. A 52 (2016) 373.
\newblock \href {http://arxiv.org/abs/1612.09432} {\path{arXiv:1612.09432}},
  \href {https://doi.org/10.1140/epja/i2016-16373-9}
  {\path{doi:10.1140/epja/i2016-16373-9}}.

\bibitem{2016PhRvC..94d4303B}
T.~{Baba}, M.~{Kimura}, {Structure and decay pattern of the linear-chain state
  in $^{14}$C}, Phys. Rev. C 94~(4) (2016) 044303.
\newblock \href {http://arxiv.org/abs/1605.05567} {\path{arXiv:1605.05567}},
  \href {https://doi.org/10.1103/PhysRevC.94.044303}
  {\path{doi:10.1103/PhysRevC.94.044303}}.

\bibitem{UJP2016}
B.~E. {Grinyuk}, D.~V. {Piatnytskyi}, {Structure of $^{14}$C and $^{14}$O
  nuclei calculated in the variational approach}, Ukr. J. Phys. 61~(8) (2016)
  674--680.

\bibitem{2017JPhCS.863a2033B}
T.~{Baba}, M.~{Kimura}, {Structure and decay pattern of linear-chain states in
  neutron-rich Carbon isotopes}, J. Phys. Conf. Ser. 863 (2017) 012033.
\newblock \href {https://doi.org/10.1088/1742-6596/863/1/012033}
  {\path{doi:10.1088/1742-6596/863/1/012033}}.

\bibitem{2017PhRvC..95f4318B}
T.~{Baba}, M.~{Kimura}, {Three-body decay of linear-chain states in $^{14}$C},
  Phys. Rev. C 95~(6) (2017) 064318.
\newblock \href {http://arxiv.org/abs/1702.04874} {\path{arXiv:1702.04874}},
  \href {https://doi.org/10.1103/PhysRevC.95.064318}
  {\path{doi:10.1103/PhysRevC.95.064318}}.

\bibitem{UJP2017}
B.~E. {Grinyuk}, D.~V. {Piatnytskyi}, {Structure of $^{14}$N nucleus within a
  five-cluster model}, Ukr. J. Phys. 62~(10) (2017) 835--844.

\bibitem{1993NuPhA.564....1T}
D.~R. {Tilley}, H.~R. {Weller}, C.~M. {Cheves}, {Energy levels of light nuclei
  \makebox{$A$} = 16-17}, Nucl. Phys. A 564 (1993) 1--183.
\newblock \href {https://doi.org/10.1016/0375-9474(93)90073-7}
  {\path{doi:10.1016/0375-9474(93)90073-7}}.

\bibitem{2014PhRvC..90e4324F}
M.~{Freer}, J.~D. {Malcolm}, N.~L. {Achouri}, N.~I. {Ashwood}, D.~W.
  {Bardayan}, S.~M. {Brown}, W.~N. {Catford}, K.~A. {Chipps}, J.~{Cizewski},
  N.~{Curtis}, K.~L. {Jones}, T.~{Munoz-Britton}, S.~D. {Pain}, N.~{Soi{\'c}},
  C.~{Wheldon}, G.~L. {Wilson}, V.~A. {Ziman}, {Resonances in $^{14}$C observed
  in the $^{4}$He($^{10}$Be ,{$\alpha$} )$^{10}$Be reaction}, Phys. Rev. C
  90~(5) (2014) 054324.
\newblock \href {https://doi.org/10.1103/PhysRevC.90.054324}
  {\path{doi:10.1103/PhysRevC.90.054324}}.

\bibitem{2016ChPhC..40k1001T}
Z.~Y. {Tian}, Y.~L. {Ye}, Z.~H. {Li}, C.~J. {Lin}, Q.~T. {Li}, Y.~C. {Ge},
  J.~L. {Lou}, W.~{Jiang}, J.~{Li}, Z.~H. {Yang}, J.~{Feng}, P.~J. {Li},
  J.~{Chen}, Q.~{Liu}, H.~L. {Zang}, B.~{Yang}, Y.~{Zhang}, Z.~Q. {Chen},
  Y.~{Liu}, X.~H. {Sun}, J.~{Ma}, H.~M. {Jia}, X.~X. {Xu}, L.~{Yang}, N.~R.
  {Ma}, L.~J. {Sun}, {Cluster decay of the high-lying excited states in
  $^{14}$C}, Chinese Physics C 40~(11) (2016) 111001.
\newblock \href {http://arxiv.org/abs/1607.00157} {\path{arXiv:1607.00157}},
  \href {https://doi.org/10.1088/1674-1137/40/11/111001}
  {\path{doi:10.1088/1674-1137/40/11/111001}}.

\bibitem{2016PhRvC..93d4323B}
S.~{Bedoor}, A.~H. {Wuosmaa}, M.~{Albers}, M.~{Alcorta}, S.~{Almaraz-Calderon},
  B.~B. {Back}, P.~F. {Bertone}, C.~M. {Deibel}, C.~R. {Hoffman}, J.~C.
  {Lighthall}, S.~T. {Marley}, D.~G. {Mcneel}, R.~C. {Pardo}, K.~E. {Rehm},
  J.~P. {Schiffer}, D.~V. {Shetty}, {Structure of $^{14}$C and $^{14}$B from
  the $^{14,15}$C(d,$^{3}$He)$^{13,14}$B reactions}, Phys. Rev. C 93~(4) (2016)
  044323.
\newblock \href {https://doi.org/10.1103/PhysRevC.93.044323}
  {\path{doi:10.1103/PhysRevC.93.044323}}.

\bibitem{2016PhRvC..93a4321F}
A.~{Fritsch}, S.~{Beceiro-Novo}, D.~{Suzuki}, W.~{Mittig}, J.~J. {Kolata},
  T.~{Ahn}, D.~{Bazin}, F.~D. {Becchetti}, B.~{Bucher}, Z.~{Chajecki},
  X.~{Fang}, M.~{Febbraro}, A.~M. {Howard}, Y.~{Kanada-En'yo}, W.~G. {Lynch},
  A.~J. {Mitchell}, M.~{Ojaruega}, A.~M. {Rogers}, A.~{Shore}, T.~{Suhara},
  X.~D. {Tang}, R.~{Torres-Isea}, H.~{Wang}, {One-dimensionality in atomic
  nuclei: A candidate for linear-chain {$\alpha$} clustering in $^{14}$C},
  Phys. Rev. C 93~(1) (2016) 014321.
\newblock \href {https://doi.org/10.1103/PhysRevC.93.014321}
  {\path{doi:10.1103/PhysRevC.93.014321}}.

\bibitem{2017NCimC..39..372S}
D.~{Suzuki}, T.~{Ahn}, D.~{Bazin}, F.~D. {Becchetti}, S.~{Beceiro-Novo},
  A.~{Fritsch}, J.~J. {Kolata}, W.~{Mittig}, {AT-TPC Collaboration}, {Cluster
  structure of neutron-rich $^{10}$Be and $^{14}$C via resonant alpha
  scattering}, Nuovo Cimento C Geophysics Space Physics C 39 (2017) 372.
\newblock \href {https://doi.org/10.1393/ncc/i2016-16372-0}
  {\path{doi:10.1393/ncc/i2016-16372-0}}.

\bibitem{2017PhLB..766...11Y}
H.~{Yamaguchi}, D.~{Kahl}, S.~{Hayakawa}, Y.~{Sakaguchi}, K.~{Abe}, T.~{Nakao},
  T.~{Suhara}, N.~{Iwasa}, A.~{Kim}, D.~H. {Kim}, S.~M. {Cha}, M.~S. {Kwag},
  J.~H. {Lee}, E.~J. {Lee}, K.~Y. {Chae}, Y.~{Wakabayashi}, N.~{Imai},
  N.~{Kitamura}, P.~{Lee}, J.~Y. {Moon}, K.~B. {Lee}, C.~{Akers}, H.~S. {Jung},
  N.~N. {Duy}, L.~H. {Khiem}, C.~S. {Lee}, {Experimental investigation of a
  linear-chain structure in the nucleus $^{14}$C}, Phys. Lett. B 766 (2017)
  11--16.
\newblock \href {http://arxiv.org/abs/1610.06296} {\path{arXiv:1610.06296}},
  \href {https://doi.org/10.1016/j.physletb.2016.12.050}
  {\path{doi:10.1016/j.physletb.2016.12.050}}.

\bibitem{2018JPhCS.966a2040D}
A.~{Di Pietro}, J.~P. {Fern{\'a}ndez-Garc{\'{\i}}a}, F.~{Ferrera},
  P.~{Figuera}, M.~{Fisichella}, M.~{Lattuada}, S.~{Marletta}, C.~{Marchetta},
  D.~{Torresi}, M.~{Alcorta}, M.~J.~G. {Borge}, T.~{Davinson}, S.~{Heinitz},
  A.~M. {Laird}, A.~C. {Shotter}, D.~{Schumann}, N.~{Soic}, O.~{Tengblad},
  M.~{Zadro}, {Experimental investigation of exotic clustering in $^{13}$B and
  $^{14}$C using the resonance scattering method}, J. Phys. Conf. Ser. 966
  (2018) 012040.
\newblock \href {https://doi.org/10.1088/1742-6596/966/1/012040}
  {\path{doi:10.1088/1742-6596/966/1/012040}}.

\bibitem{2009PAN....72....6G}
B.~E. {Grinyuk}, I.~V. {Simenog}, {Structure of the $^{6}$He nucleus in the
  three-particle model}, Phys. Atom. Nucl. 72 (2009) 6--19.
\newblock \href {https://doi.org/10.1134/S1063778809010025}
  {\path{doi:10.1134/S1063778809010025}}.

\bibitem{2014PAN....77..415G}
B.~E. {Grinyuk}, I.~V. {Simenog}, {Structural properties of the $^{10}$Be and
  $^{10}$C four-cluster nuclei}, Phys. Atom. Nucl. 77 (2014) 415--423.
\newblock \href {https://doi.org/10.1134/S1063778814030090}
  {\path{doi:10.1134/S1063778814030090}}.

\bibitem{2011NPAE}
B.~E. {Grinyuk}, I.~V. {Simenog}, {Structure Peculiarities of Three- and
  Four-Cluster Nuclei $^{6}$He, $^{6}$Li, and $^{10}$Be, $^{10}$C}, Nucl. Phys.
  Atom. En. 12~(1) (2011) 7--15.

\bibitem{1971PhLB...34..581N}
V.~G. {Neudatchin}, V.~I. {Kukulin}, V.~L. {Korotkikh}, V.~P. {Korennoy}, {A
  microscopically substantiated local optical potential for
  {\ensuremath{\alpha}}- {\ensuremath{\alpha}}- scattering}, Phys. Lett. B
  34~(7) (1971) 581--583.
\newblock \href {https://doi.org/10.1016/0370-2693(71)90142-0}
  {\path{doi:10.1016/0370-2693(71)90142-0}}.

\bibitem{ElChYa1979}
V.~I. {Kukulin}, V.~G. {Neudatchin}, Y.~F. {Smirnov}, {An interaction of
  compound particles and the Pauli principle}, Sov. J. Part. Nucl. 10 (1979)
  1006.

\bibitem{UJP2007}
B.~E. {Grinyuk}, D.~V. {Piatnytskyi}, I.~V. {Simenog}, {Structure
  Characteristics of a $^{4}$He Nucleus within the Microscopic Approach}, Ukr.
  J. Phys. 52~(5) (2007) 424--435.

\bibitem{1977JPhG....3..795K}
V.~I. {Kukulin}, V.~M. {Krasnopol'sky}, {A stochastic variational method for
  Few-Body Syst.}, J. Phys. G Nucl. Phys. 3 (1977) 795--811.
\newblock \href {https://doi.org/10.1088/0305-4616/3/6/011}
  {\path{doi:10.1088/0305-4616/3/6/011}}.

\bibitem{SuzukiVarga}
{Y.~{Suzuki}, K.~{Varga}}, Stochastic Variational Approach to Quantum
  Mechanical Few-Body Problems, Springer-Verlag, Berlin, Heidelberg, 1998.

\bibitem{2003PrPNP..51..223H}
E.~{Hiyama}, Y.~{Kino}, M.~{Kamimura}, {Gaussian expansion method for few-body
  systems}, Prog. Part. Nucl. Phys. 51~(1) (2003) 223--307.
\newblock \href {https://doi.org/10.1016/S0146-6410(03)90015-9}
  {\path{doi:10.1016/S0146-6410(03)90015-9}}.

\bibitem{2018FrPhy..13.2106H}
E.~{Hiyama}, M.~{Kamimura}, {Study of various few-body systems using Gaussian
  expansion method (GEM)}, Frontiers of Physics 13~(6) (2018) 132106.
\newblock \href {http://arxiv.org/abs/1809.02619} {\path{arXiv:1809.02619}},
  \href {https://doi.org/10.1007/s11467-018-0828-5}
  {\path{doi:10.1007/s11467-018-0828-5}}.

\bibitem{1956PhRv..104.1466H}
R.~H. {Helm}, {Inelastic and Elastic Scattering of 187-Mev Electrons from
  Selected Even-Even Nuclei}, Phys Rev. 104 (1956) 1466--1475.
\newblock \href {https://doi.org/10.1103/PhysRev.104.1466}
  {\path{doi:10.1103/PhysRev.104.1466}}.

\bibitem{1982NuPhA.379..523S}
L.~A. {Schaller}, L.~{Schellenberg}, T.~Q. {Phan}, G.~{Piller}, A.~{Ruetschi},
  H.~{Schneuwly}, {Nuclear charge radii of the carbon isotopes $^{12}$C,
  $^{13}$C and $^{14}$C}, Nucl. Phys. A 379~(3) (1982) 523--535.
\newblock \href {https://doi.org/10.1016/0375-9474(82)90012-4}
  {\path{doi:10.1016/0375-9474(82)90012-4}}.

\bibitem{2013ADNDT..99...69A}
I.~{Angeli}, K.~P. {Marinova}, {Table of experimental nuclear ground state
  charge radii: An update}, Atomic Data and Nuclear Data Tables 99~(1) (2013)
  69--95.
\newblock \href {https://doi.org/10.1016/j.adt.2011.12.006}
  {\path{doi:10.1016/j.adt.2011.12.006}}.

\bibitem{1967PhRv..160..874F}
R.~F. {Frosch}, J.~S. {McCarthy}, R.~E. {Rand}, M.~R. {Yearian}, {Structure of
  the He$^{4}$ Nucleus from Elastic Electron Scattering}, Phys. Rev. 160~(4)
  (1967) 874--879.
\newblock \href {https://doi.org/10.1103/PhysRev.160.874}
  {\path{doi:10.1103/PhysRev.160.874}}.

\bibitem{1992PhRvL..68.3841B}
P.~E. {Bosted}, L.~{Andivahis}, A.~{Lung}, L.~M. {Stuart}, J.~{Alster}, R.~G.
  {Arnold}, C.~C. {Chang}, F.~S. {Dietrich}, W.~{Dodge}, R.~{Gearhart},
  J.~{Gomez}, K.~A. {Griffioen}, R.~S. {Hicks}, C.~E. {Hyde-Wright},
  C.~{Keppel}, S.~E. {Kuhn}, J.~{Lichtenstadt}, R.~A. {Miskimen}, G.~A.
  {Peterson}, G.~G. {Petratos}, S.~E. {Rock}, S.~{Rokni}, W.~K. {Sakumoto},
  M.~{Spengos}, K.~{Swartz}, Z.~{Szalata}, L.~H. {Tao}, {Measurements of the
  electric and magnetic form factors of the proton from Q$^{2}$=1.75 to 8.83
  (GeV/c)$^{2}$}, Phys. Rev. Lett. 68~(26) (1992) 3841--3844.
\newblock \href {https://doi.org/10.1103/PhysRevLett.68.3841}
  {\path{doi:10.1103/PhysRevLett.68.3841}}.

\end{thebibliography}

\end{document}